\documentstyle[12pt]{article}
\begin{document}
\title{
THE $q$-DEFORMED WIGNER OSCILLATOR
IN QUANTUM MECHANICS}       
\author{R. de Lima Rodrigues\footnote{Permanent address: Departamento de
Ci\^encias Exatas e da Natureza, Universidade Federal de Campina Grande,
Cajazeiras - PB, 58.900-000 - Brazil. 
E-mail to RLR is rafaelr@cbpf.br or
rafael@fisica.ufpb.br. }
\\ Centro Brasileiro de Pesquisas F\'\i sicas (CBPF)\\
Rua Dr. Xavier Sigaud, 150,CEP 22290-180,\\
 Rio de Janeiro, RJ, Brazil
}
\date{ }
\maketitle

\begin{abstract}
Using a super-realization of the Wigner-Heisenberg
algebra a new realization of the $q$-deformed 
Wigner oscillator is implemented.
\end{abstract}

\vspace{0.5cm}
PACS numbers: 03.70.+k, 11.10.Ef

\vspace{2cm}

Dedicated to the memory of Prof. Jambunatha Jayaraman, 
28 January 1945-19 June 2003.

\newpage
\section{Introduction}

\paragraph*{}

In 1989, independently, Biedenharn and Macfarlame \cite{mac89}, 
introduced the $q$-deformed harmonic oscillator and constructed
a realization of the $SU_q(2)$ algebra, using a $q$-analogue of the
harmonic oscillator and the Jordan-Schwinger mapping.
The $q$-deformation of $SU(2),$ denoted by $SU_q(2),$ is one of the 
simplest examples of a quantum group.

The deformation of the conventional quantum mechanical laws
has been implemented via different definitions  and studied by several 
authors in the literature \cite{WH,Kuli90,Cutri90,macfa93,Cho94,Mik97,Bonatsos,palev03}.
Also, recently Palev {\it et al.} have investigated 
the 3D Wigner oscillator \cite{palev03}.

The main purpose of this work is to set up a realization of the $q$-deformed 
Wigner oscillator \cite{WH}.

\section{The $q$-deformed usual harmonic oscillator}

\paragraph*{}

In this section, we consider
the $q$-deformed ladder operators of the harmonic oscillator, 
$a^-$ and its adjoint $a^+,$ 
acting on the basis $\mid n>, \quad n= 0, 1, 2, \cdots,$ as
\cite{mac89}
$
a^-_q\mid 0>= 0, \quad \mid n>= \frac{(a^+_q)^n}{([n]!)^{\frac 12}}\mid 0>
$
where
$
[n]!= [n][n-1]\cdots[1].
$
The classical limit $q\rightarrow 1$ yields to the conventional ladder boson
operators $a^{\pm},$  which satisfies 
$[a^-, a^+]=1, \quad 
a^-|n>=\sqrt{n}|n-1>,\quad a^+|n>=\sqrt{n+1}|n+1>.
$

On the other hand,
$su(1,1)$ algebra satisfies the following commutation relations
$[K_0, K_{\pm}]= \pm K_{\pm}, \quad [K_+, K_-]= -2K_0
$
and the Casimir operator is given as
$C= K_0(K_0 - 1)-K_+K_-, 
$
where $K_0|0>=k_0|0>$ and $K_-|0>=0.$
A usual representation for this algebra is given in terms of the ladder 
operators $a^-= (x+ip)/\sqrt{2}, \quad a^+= (x-ip)/\sqrt{2}.$ 
The $su(1,1)$ generators are given as $K_0= \frac 12\left(N+\frac 12\right), 
\quad K_+= \frac 12(a^+)^2$ and $K_-= \frac 12 (a^-)^2,$ where $N= a^+a.$ 
Thus, the Casimir operator is given by $C= -\frac{3}{16}.$ This system has two 
different representations whose $k_0$ is $\frac 14$ and $\frac 34$. 

Its $q$-deformation, $su_{q^2}(1,1),$ is given \cite{Kuli90} as

\begin{equation}
\label{c3}
[\tilde{K}_0, \tilde{K}_{\pm}]= \pm \tilde{K}_{\pm} \quad 
[\tilde{K}_+, \tilde{K}_-]= -[2\tilde{K}_0]_{q^2}, \quad 
[x]_{\mu}\equiv (\mu^x - \mu^{-x})/(\mu - \mu^{-1}).
\end{equation}   
In Ref. \cite{Cutri90} was found a realization of the 
$su_{q^2}(1,1)$ in terms of the generators 
of $su(1,1)$.
The $q$-deformed ladder operators satisfy

\begin{equation}
\label{m8}
a^-_qa^+_q= [N+1], \quad
a^+_qa^-_q= [N],
\end{equation}
where $N$ is the number operator
which is positive semi-definite. 
The q-analogue operators can be found in terms of the usual ladder 
boson operators $a^-$ and $a^+$.

Note that we can write
$
\mid n>= \frac{a^+_q}{\sqrt{[n]!}}\frac{(a^+_q)^{n-1}}{([n-1]!)^{\frac 12}}\mid 0>
=\frac{a^+_q}{\sqrt{[n]!}}\mid n-1>
$ 
so that we obtain

\begin{equation}
\label{m6}
a^+_q\mid n-1>= [n]^{\frac 12}\mid n>\Rightarrow 
a^+_q\mid n>= [n+1]^{\frac 12}\mid n+1>.
\end{equation}
 Also, from (\ref{m8}) and $a^-_qa^+_q\mid n>= [n+1]^{\frac 12}a^-_q\mid n+1>$
we get

\begin{equation}
\label{m7}
a^-_q\mid n>= [n]^{\frac 12}\mid n-1>.
\end{equation}
It's easy to verify that
$[N, a^+_q]= a^+_q,\quad \quad [N, a^-_q]= -a^-_q, 
\quad [N, q^N]=[a^-_qa^+_q, q^N]=0, \quad a^-_qa^+_q - qa^+_q a^-_q= q^{-N}.
$
We will show that a structure of this type 
exists for the Wigner oscillator.

\section{The q-deformed Wigner Oscillator}

\paragraph*{}

The one-dimensional Wigner super-oscillator Hamiltonian in terms of
the Pauli's matrices ($\sigma_i,$ i=1,2,3) is given by

\begin{equation}
H(\lambda +1)= \left(
\begin{array}{cc}
H_{-}(\lambda )&0\\
0&H_{+}(\lambda )
\end{array}\right),H_{-}(\lambda ) = {1\over 2}\left\{ - {d^{2}\over dx^{2}} +
x^{2} + {1\over x^2} \lambda (\lambda +1)\right\},
\end{equation}
where $H_{+}(\lambda )=H_{-}(\lambda +1).$ 
The even sector $H_{-}(\lambda )$  is the Hamiltonian of the 
oscillator with barrier
or isotonic oscillator or Calogero interaction.

Thus, from the super-realized Wigner oscillator, its first order ladder 
operators  given by \cite{WH}
$
a^{\pm }(\lambda +1) = {1\over \sqrt{2}}\left\{\pm {d\over dx} \pm
{(\lambda +1)\over x}\sigma_{3} - x\right\}\sigma_{1},
$
the Wigner Hamiltonian 
and the Wigner-Heisenberg(WH) algebra ladder relations are readily obtained as

\begin{equation}
\label{Ew7}
H(\lambda +1) = {1\over 2}\left[a^{+}(\lambda +1), 
a^{-}(\lambda +1)\right]_{+}, \left[H(\lambda +1), 
a^{\pm }(\lambda +1)\right]_{-} =
\pm a^{\pm }(\lambda +1).
\end{equation}
Equations (\ref{Ew7}) and the commutation relation

\begin{equation}
\label{Ew8}
\left[a^{-}(\lambda +1), a^{+}(\lambda +1)\right]_{-}=
1 + 2(\lambda +1)\sigma_{3}
\end{equation}
constitutes the WH algebra \cite{WH} or deformed Heisenberg algebra
\cite{macfa93,Mik97}.

Let us consider an extension of the $q$-deformed 
harmonic oscillator commutation relation, 

\begin{equation}
\label{qw1}
a_W^-a^+_W - qa^+_W a_W^-= q^{-N}(1+c\sigma_3), \quad c=2(\lambda +1)
\end{equation}
as a $q$-deformation of the Wigner oscillator commutation realization.
These operators may be written in terms of the Wigner oscillator 
ladder operators, viz.,

\begin{equation}
a_W^-=\beta(N)a^-(\lambda +1), \qquad a_W^+=a^+(\lambda +1)\beta(N), 
\quad N=a^+(\lambda +1)a^-(\lambda +1).
\end{equation}

Acting the ladder 
operators of the WH algebra in the Fock space, spanned by the  vectors  
$$
a^-(\lambda +1)|2m>_c=\sqrt{2m}|2m-1>_c,
$$
$$
a^-(\lambda +1)|2m+1>_c=\sqrt{2m+c+1}|2m>_c,
$$
$$
a^+(\lambda +1)|2m>_c= \sqrt{2m+c+1}|2m+1>_c, 
$$
$$
a^+(\lambda +1)|2m+1>_c=\sqrt{2(m+1)}|2m+2>_c,
$$
we obtain a recursion relation given by

\begin{equation}
(2m+2-c)\beta^2(2m+1)-q(2m+1+c)\beta^2(2m)=q^{-(2m+1)}(1-c).
\end{equation}
This has, for the odd quanta and $c=0,$ the following solution 

\begin{equation}
\beta(2m+1)=
\sqrt{\frac{1}{2m+2}\frac{q^{2m+2}-q^{-(2m+2)}}{q-q^{-1}}}
\Rightarrow
\beta(N)=\sqrt{\frac{[N+1]}{N+1}}.
\end{equation}
Thus, the $q$-deformed Wigner Hamiltonian and the 
commutator $[a^-_W, a^+_W]$, for $c=0$ become the 
q-deformed harmonic oscillator

\begin{equation}
H_W=\frac 12 [a_W^-,a_W^+]_+=H_b=\frac 12([N+1]+[N]), \quad
[a^-_W, a^+_W]=[N+1]-[N].
\end{equation}
Also, from even quanta this same result 
is readily found for $c=0.$

\section{Conclusion}

\paragraph*{}

In this work, we firstly presented a brief review on the 
$q$-deformation of the conventional quantum mechanical laws
for the unidimensional harmonic oscillator.
We have also implemented a new approach for the WH
algebra. Indeed, the $q$-deformations of WH algebra
are investigated via the super-realization
introduced by Jayaraman-Rodrigues \cite{WH}.  

Also, we do not assume the relations of operators $a_W^+a_W^-$ and 
$a_W^-a_W^+$. They are derived from our defining set of  
relations $a_W^-=a_q^-=\sqrt{\frac{[N+1]}{N+1}}a^-$ and 
$a^+_W=a_q^+=a^+\sqrt{\frac{[N+1]}{N+1}},$ for vanish Wigner parameter 
$(c=0)$ given by Eq. (\ref{m8}).
The case with $c\neq0$ will be presented in a forthcoming paper.
 
\vspace{2cm}
\centerline{\bf Acknowledgments}

This research was supported in part by CNPq (Brazilian Research Agency).
RLR wish to thanks the staff of the CBPF and DCEN-CFP-UFCG for the
facilities.  The
author would also like to thank V. B. Barbosa and M. A. Rego-Monteiro
by discussions.
The author is also grateful to the organizing committee of the 
 Brazilian National Meeting on Theoretical and Computational Physics, 
April 6 to 9, 2003, Brasilia-DF, Brazil.

\end{document}